\documentclass[prl,letterpaper,twocolumn,showpacs]{revtex4-1} 
\usepackage{times,xspace} 
\usepackage{amsbsy,amssymb,amsmath,bm,bbold} 
\usepackage{graphicx,color,epsfig,rotate} 
\usepackage{fancyhdr} 
\usepackage{epstopdf}
\usepackage{float}
\usepackage[colorlinks=true,linkcolor=blue,citecolor=blue,urlcolor=blue]{hyperref}

\def\bbbc{{\mathchoice {\setbox0=\hbox{$\displaystyle\rm C$}\hbox{\hbox 
to0pt{\kern0.4\wd0\vrule height0.9\ht0\hss}\box0}} 
{\setbox0=\hbox{$\textstyle\rm C$}\hbox{\hbox 
to0pt{\kern0.4\wd0\vrule height0.9\ht0\hss}\box0}} 
{\setbox0=\hbox{$\scriptstyle\rm C$}\hbox{\hbox 
to0pt{\kern0.4\wd0\vrule height0.9\ht0\hss}\box0}} 
{\setbox0=\hbox{$\scriptscriptstyle\rm C$}\hbox{\hbox 
to0pt{\kern0.4\wd0\vrule height0.9\ht0\hss}\box0}}}}

\begin{document} 
%\title{Nonreciprocal Optical Isolation in a Dual-mode Planar Waveguide Exhibiting Exceptional Points}
%\title{Asymmetric mode conversion around exceptional points in a Quasi-Guided Dual-Mode Planar Waveguide: Toward nonreciprocal optical isolation}
\title{Ultra-sensitive Light Confinement Driven by Multiple Bound States in the Continuum}
\author{ Harsh K. Gandhi,$^1$ Arnab Laha,$^{1,2}$ and Somnath Ghosh$^{1,}$}
\email{somiit@rediffmail.com}
\affiliation{$^1$Department of Physics, Indian Institute of Technology Jodhpur, Rajasthan-342037, India\\
	$^2$ Institute of Radio Physics and Electronics, University of Calcutta, Kolkata 700009, India}
	
\begin{abstract} 

 We propose a unique framework to study the topological properties of an optical bound state in the continuum(BIC). We employ the interactions between proximity resonances undergoing avoided resonance crossing in a specialty optical microcavity. We utilize a continuous system parameter tuning to induce destructive interference between the resonances with cancelling leakage losses. Similar to the physical insight of Friedrich-Wingten type-BIC, we demonstrate the evolution of an ultra-high quality mode. We report the formation of a special-BIC line in the system parameter space connecting locations of multiple quasi-BICs. Aiming to develop a novel scheme to enhance the performance of optical sensing in microcavity, we study the sensitivity of transmission coefficients and quality factor to sense even ultra-small perturbations in the system configuration. 
 
\end{abstract} 
 
%\pacs{03.65.Yz, 02.40.Xx, 03.65.Vf, 02.10.Yn, 42.25.-p, 42.82.-m} 
 
\maketitle % 

Physics governing the interaction between resonances in optical-platforms has become an ever-increasing area of research and development in the past few years. With particular emphasis on the modeling photonic systems such as optical microcavities towards desired optical performance has been extensively studied \cite{Vahala}. Optical microcavities have the ability to confine light, enabling the formation of the high-energy-density of light. This phenomenon is essential for various applications that are prevalent in the field of nonlinear optics \cite{Lin}, biosensing \cite{biosensing}, low threshold lasing \cite{Lasing}, processing quantum informatics \cite{Quantum}, to name a few. However, their performance is impaired by the task of identifying long-confined resonances with lower power-requirements. Unlike bulk-media \cite{Soljacic}, it is possible to enhance the Purcell factor by increasing the $Q/V$ ratio, which is the effective quality-factor ($Q$) of light over the modal volume \cite{yuri}. In this context, super-cavity resonances have shown promise to accomplish the same \cite{Tolstoy,bic2,bic,bic3,Hsu1}. Recently, a novel approach involving interaction between two proximity resonances via avoided resonance crossing (ARC) phenomenon has been identified to hold great potential for the evolution of high-$Q$ modes \cite{Harsh1,Laha}. Furthermore, the application of the concept of ARC has been widely demonstrated in various resonator structures such as rectangular, semistadium, and elliptical microcavities \cite{ARC1}.  
%Recently, different approaches like interband photonic transition \cite{Yu09}, nonlinear resonance shifts \cite{Yu15} have reported on chip-scale devices, however, the foremost barrier is requirement of high threshold power for operation.

In particular, the formation of one such high-$Q$ states being the quasi-Bound states in the continuum (BIC) \cite{Song}, has been extensively studied and reported.  BICs are mathematical abstractions wherein a photonic BIC is a state that captures high-volume of energy, with theoretically infinite efficiency with closed radiation channels despite being in the continuum band \cite{Stillinger}. In the case of a cavity, a BIC would have no leakage radiation outside the cavity and would theoretically confine light for an infinite time. However, a true BIC would require infinite-size of the cavity or an infinitely high permittivity of the medium \cite{BIC}, which is, however, not physically implementable. Further, the practical feasibility of the realization of BIC in various optical platforms is still an open area of research \cite{Marinica,Ordonez,Carletti}.  In this direction, it has been demonstrated that two cavity-supported resonances under a strong-coupling regime undergoing destructive interference, could lead to one of them evolving with a higher $Q$ \cite{bic, bic2, bic3, Hsu1}. This is in agreement with the Friedrich-Wingten theory of continuous parameter tuning \cite{FW}, leading to destructive interference patterns \cite{jin} between the resonances canceling out the losses from the radiation channel of the lossy modes, thus attaining an impulsive growth in quality factor. The evolution of such finite high-$Q$ resonances is termed as quasi-BICs.

In this letter, we explore the topological properties of such high-$Q$ quasi-BICs. We present a scheme where we show the presence of spatial variation of gain-loss in an optical resonator and obtain enhancement in $Q$ up to four orders of magnitude even in an overall lossy system. Devoid the interference between Mie resonances, we show the formation of a quasi-BIC. We show the possibility of carefully tuning and optimizing the material gain-loss profile to sustain a quasi-BIC. We employ ARC to show that the enhancement in quality factor is a consequence of quasi-BIC state. Furthermore, we show, for the first time to the best of our knowledge, the formation of a special-BIC line in the parameter space connecting multiple such quasi-BICs. This special-BIC line provides a new degree of freedom to explore further topological aspects of the BIC and also physically support any cascading nonlinear effects \cite{Carletti2}. Aiming to develop a novel scheme to enhance optical sensing in microcavity, and we show the ultra-high sensitivity of transmission coefficients and quality of light state to be able to sense even slight perturbations.

In an ideal scenario, a resonance would exhibit nearly infinite lifetime, in a case where the material loss is negligible. However, a practical scenario would contain material absorption losses that need to be dealt with. Therefore, the interaction and interference between proximity resonances could be extensively understood using non-Hermitian quantum mechanics. In this case, when we consider the interaction between two proximity resonances, the system could be expressed as:
\begin{equation}
H=\begin{bmatrix}
\eta_1 & 0\\
0 &  \eta_2 \\
\end{bmatrix}+
\begin{bmatrix}
0 & V\\
W & 0 \\
\end{bmatrix}=\begin{bmatrix}
\eta_1 & V\\
W &  \eta_2 \\
\end{bmatrix}
\end{equation}
where $\eta_{1,2}$ are the complex passive states of the system, which are subjected to an external off-diagonal perturbation $(V,W)$ similar to the system shown in \cite{Harsh1}. The formation of the BIC using more than two proximity resonances would also follow a similar approach. However, we limit our discussion to exploring the features of a quasi-BIC from two proximity resonances. To ensure we limit the external coupling to 0, we impose the condition of $V$=$W^*$. Under such constraint, we can calculate the energy eigen-values of the system as $E_{1,2}=({\eta_1+\eta_2})/{2} \pm \sqrt{{(({\eta_1-\eta_2})/{2})}^2+|V|^2}$. Without any loss in generality, it is safe to say that on the basis of the value of $V$, the resonances undergo ARC under different coupling regimes \cite{Teller}. This ARC thus established generally induces mutual energy exchange, which in particular evolves with a longer-living state along with a subsequent decay in the lifetime of the latter \cite{bic,bic2}.  Note that the system mentioned above is mathematically realized, but for the sake of physical implementation, we exploit the correlation between the poles of the scattering matrix ($S$-matrix) and the eigenvalues of the hamiltonian. This physical equivalence has been exploited to a great extent in \cite{Laha,Harsh1}. 
\begin{figure}[b]
	\centering
	\includegraphics[width=\linewidth]{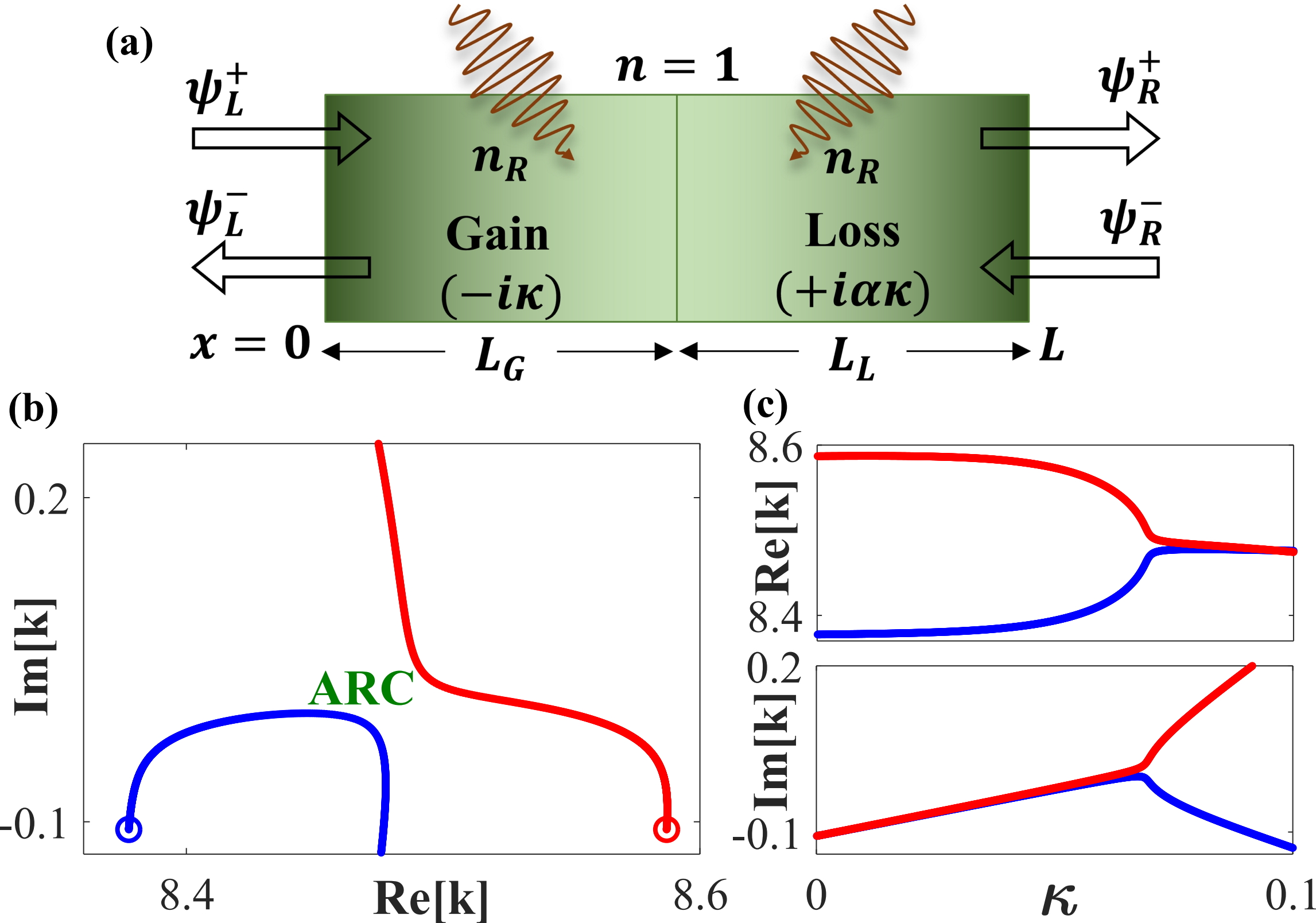}
	\caption{{\bf(a)} Schematic of gain-loss assisted 1$D$ Fabry-P\'erot type microcavity occupying the region $0\le x\le L$. $\{\psi_L^{+},\psi_{L}^-\}$ and $\{\psi_R^{+}, \psi_{R}^-\}$ represent the complex incident and scattered wavefunctions respectively. {\bf(b)} The dynamics of a pair of chosen coupled-poles of $S$-matrix showing ARC in complex $k$-plane with an increasing $\kappa$ for a fixed $\alpha=2.23$. {\bf(c)} The corresponding variation of the Re[$k$] (upper panel) and Im[$k$] (lower panel) with $\kappa$.}
	\label{p2}
\end{figure}
Using this knowledge, we design a two-port 1D Fabry-Perot type optical microcavity, in which we carefully select spatially varying complex gain-loss profile in a constant real background refractive index $n_R$ as can be seen in Fig 1(a). We give special emphasis on integrated photonic devices and hence for this study we restrict the value of real refractive index $n_R=1.5$ for silica ($SiO_2$).  Note that there is no restriction in the selection of the geometry of the cavity, nor is there any constraint in the dimensions of the structure. We choose Fabry-Perot type resonators for the sake of understanding interference physics along the axis of the cavity. Along the length of the resonator $L=10\mu m$, we divide into two sections having lengths $L_G$ and $L_L$, which are the lengths of the material gain and material loss dominated regions. For the sake of simplicity, we take equal lengths for both sections, $L_L=L_G=5 \mu m$. The refractive indices are defined as $n_G=n_R-i\kappa$ and $n_L=n_R+i\alpha \kappa$, where $n_G$ and $n_L$ are the complex  refractive indices of the gain and loss regions. Here $\kappa$ is the gain-coefficient, and $\alpha$ is the ratio of loss to gain coefficients. The introduction of $\kappa$ serves the purpose of introducing the system's non-hermiticity. It also provides the ability to tune the system material for enhanced feasibility for fabrication. This tuning ability is per the causality condition of the Kramer-Konig relationship, which permits the flexibility of independent tuning of $Im[n]$ at a single frequency of operation \cite{Phang}. A further selection of the virtual states represented by Eq.1 is made by calculating the poles of the $S$-matrix, which is of the form of:
\begin{equation}
\begin{bmatrix}
B\\
C
\end{bmatrix}
= S(n(x),\omega)\begin{bmatrix}
A\\
D
\end{bmatrix}
\end{equation}
where the incident and the scattered complex wavefunctions are represented by $A,D$ and $B,C$ respectively. This calculation of the poles of $S$=matrix is numerically executed with the help of $1/max(Eig(S))=0$, where $max(Eig(S))$ is the maximal-modulus eigenvalues of the $S$-matrix. As mentioned above, the entire analysis revolves around the fact that the poles obtained are indicative of the scattering states present inside the cavity. These proximity resonances contain the information related to the continuous confinement of the particle responsible for the scattering of the light and hence is justified to be equivalent to the poles of the S-matrix in the complex frequency plane ($k$-plane). For a given complex pole in the $k$-plane, the real part is indicative of the eigen-frequencies at which the linear scattering is exhibited by the system. However, the imaginary part would depict the coupling loss that the pole undergoes during ARC. Here, it becomes vital to state that the coupling loss, which is the channel for radiation of the quasi-BIC, must be limited to null values for it to exhibit diverging lifetime. The advent of the introduction of gain-loss within the system serves the purpose of reducing this coupling loss to theoretically zero values, manifesting nearly no leakage radiation from the quasi-BIC resonance. 

 We demonstrate the same, using two proximity resonances chosen in accordance with the structure dimensions as per \cite{Ge}. We choose two proximity resonances between 8.3 and 8.7 $\mu m$ in the $Re[k]$. With the introduction of gain-loss, these proximity resonances now under the coupling regime governed by the external perturbation are forced to interact mutually. With a fixed parameter $\alpha$=2.23, we increase the gain from 0 to 0.1. Upon this exercise, we observe the occurrence of an ARC phenomenon, with $\Re[k]$ undergoing crossing and the $\Im[k]$ having an anti-cross as can be seen in fig 1(b),(c). 
Upon close inspection, we observe that the point of ARC in the system when the $Re[k]$ approaches 8.475 $\mu m$. The selection of parameters for $\alpha$ was on the basis of the observation of the divergence of the quality factor, which is defined as $0.5Re[k]/Im[k]$ where $k$ is the complex pole of the $S$-matrix corresponding to the proximity resonances. Furthermore, we observe a divergence in $Q$ of one of the resonances at a particular value of $\kappa=0.0527$, where the $Q$ diverges to more than four orders of magnitude as can be seen in fig 2(a). In contrast, subsequent decay in the other resonance is also observed. The $Q$-factor of the pole at the quasi-BIC state, we denote as $Q_{BIC}\approx 7.78$ x $10^5$, while the subsequent decaying mode had a $Q$-factor $Q_{decay} \approx 7$. This simultaneous decay of one mode and the divergence is of much importance for the understanding of why we call the diverging resonance as a quasi-BIC. For a given carefully selected parameters of the system, we observe that one of the resonances decays in lifetime having characteristics similar to a continuum mode with very high leakage radiation, while the other despite being in the same system preset, undergoes enhanced lifetime from very low to ultra-high $Q$-factor.

\begin{figure}[ht]
	\centering
	\includegraphics[width=\linewidth]{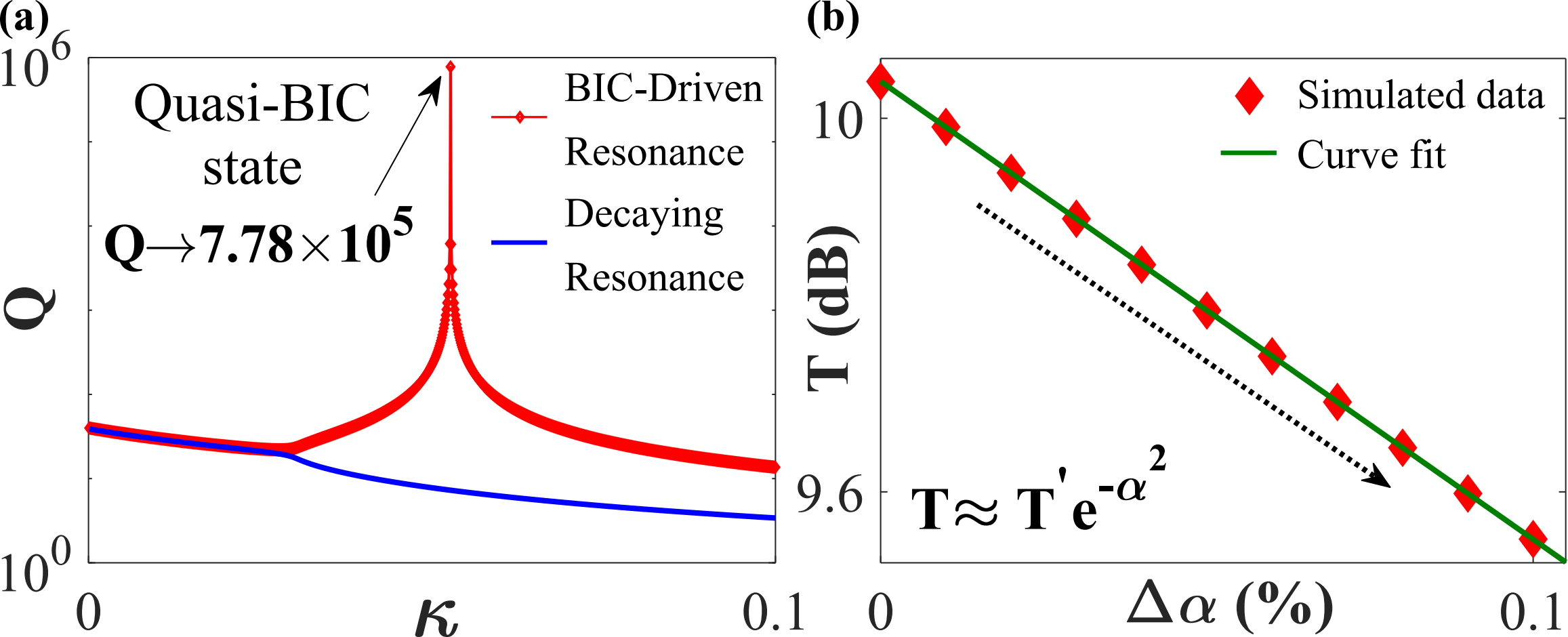}
	\caption{{\bf(a)} $Q$-factor of the chosen coupled poles as a function of $\kappa$. Here, the divergence of the $Q$-factor of a specific pole (represented by red-line with diamond marker), has been observed near $\kappa=0.0527$ which indicates the quasi-BIC state. The $Q$-factor has been enhanced by more than 4 orders of magnitude than the decaying pole (represented by blue line). {\bf(b)} Transmission sensitivity due to the slight variation in $\alpha$ (in terms of $\Delta\alpha$) from quasi-BIC state. The point at $\Delta\alpha=0$ represent the transmission coefficient of quasi-BIC state as indicated in (a), where the other points indicates different transmission coefficients for different $\Delta\alpha$.} 
	\label{p2}
\end{figure}

Another important aspect of our system is that it is overall lossy. We ensure this by taking the value of $\alpha$ significantly higher than 1. This is justified for our preset where we take the spatial dimensions of the gain and loss regions as equal. This is of very high physical significance, as natural materials that are considered always posses material absorption losses. Therefore, the amount of gain, which is lower than the amount of loss, has the function of inducing interference patters that are out of phase and hence destructive in nature.
\begin{figure}[b]
	\centering
	\includegraphics[width=\linewidth]{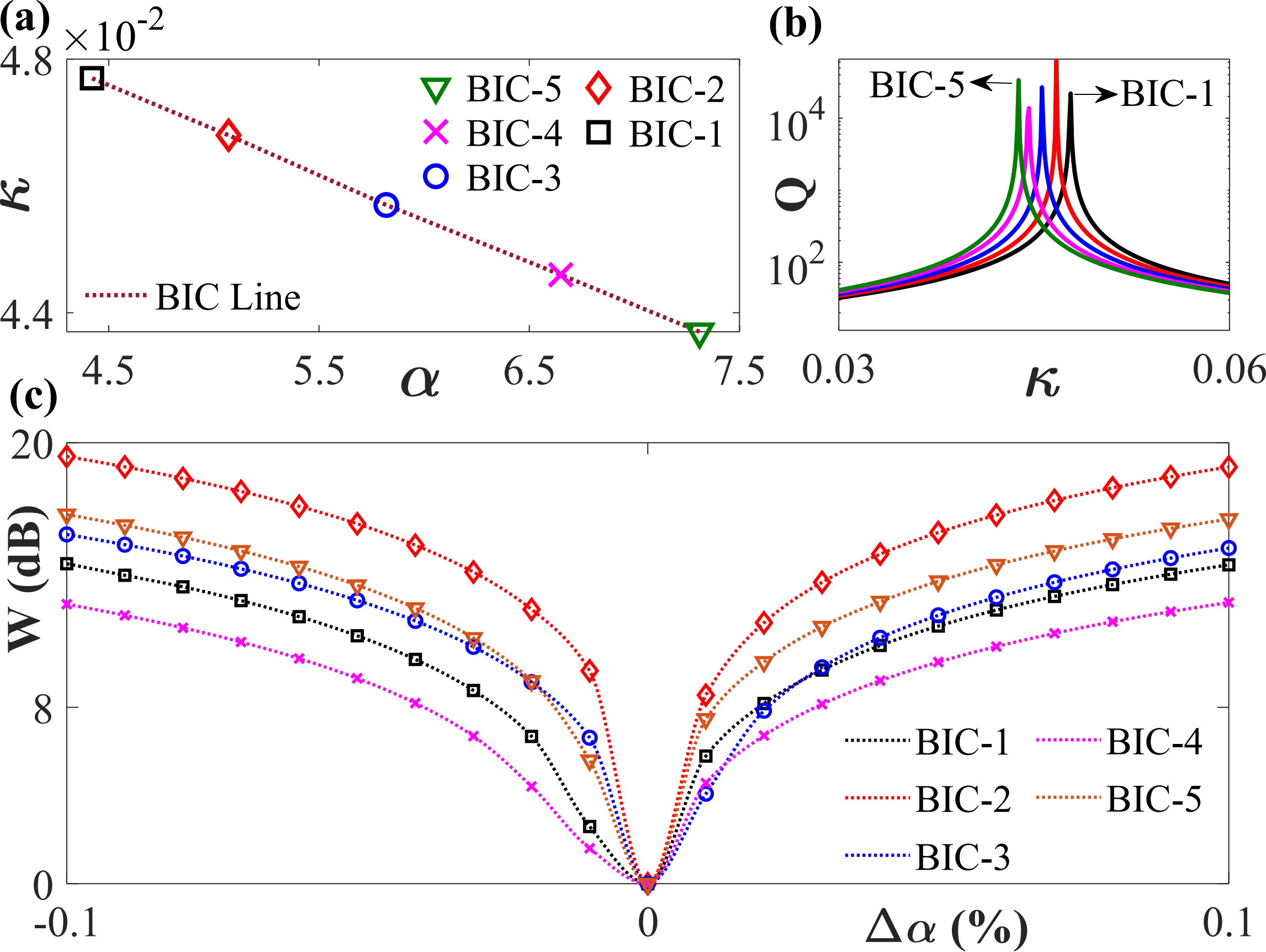}
	\caption{{\bf(a)} Multiple quasi-BIC states (indicated by figure legends) between the same pair of coupled poles for different gain-loss pumping in $(\kappa,\alpha)$-plane; which are forming a straight line (fitted), coined as special BIC-line (indicated by dotted brown line). {\bf(b)} The verification of these five quasi-BICs in terms of diverging $Q$-factors. The different colors of $Q$-factors of five quasi-BICs correspond to the colors used to indicate the respective quasi-BIC points in (a). {\bf(c)} Sensitivity coefficients $W$ with the variation of $\Delta\alpha$ due to slight perturbation for all the five quasi-BICs. All of them show at least 3.4 dB sensitivity response for $1\%$ change in the $\alpha$.} 
	\label{p2}
\end{figure}
The extremely diverging values of $Q$ we obtain from a quasi-BIC driven resonance has its dependence on a variety of system parameter. In fact, the quasi-BIC could also be explored with the change in the length of both the gain and loss regions. However, in this letter, we limit our discussion by fixing our physical dimensions as constants so as to realize the effects of other critical system parameters, namely the ones that govern the system absorption losses. In our system, these parameters are $\kappa$ and $\alpha$. For this purpose, to investigate the potential sensing abilities that these structures provide, we try to introduce an irregularity in the gain-loss profile. However, to carry this out in a systematic manner, we keep the material of the gain region intact and vary the material of the second block that is loss dominated. This can be done with the introduction of small index-modifications that can change the loss coefficient. We only introduce index-change as low as 1\% variation in $\alpha$ so as to test the sensitivity for small perturbations. In summary, we vary the $\alpha$ for a fixed gain. We further study the transmission coefficients of the system under parameters that lead to the quasi-BIC and at various different parameter presets near the quasi-BIC in the system parameter space. As can be seen in figure 2(b), we vary the parameter $\alpha$, keeping the gain pumped as constant. Here for $\Delta \alpha=0$, we get transmission coefficients at quasi-BIC state. Now we find the transmission of light through the structure for all different systems that we derive by changing $\alpha$ by 1\% to 10\% of the original $\alpha=2.23$. By plotting them with the change in $\alpha$ in log scale, and subsequently curve fitting them, we observe an inverse square dependence of loss to gain ratio on transmission. We find that $log(T)\approx log(T_0) -\alpha^2$, where $T_0$ refers to the transmission coefficient of the system at the quasi-BIC parameter configuration. 

Utilizing the same proximity resonances, we work out different configurations of the system with constant length, where there is enhanced trapping of light. In this context, we confirm the claim with the help of ARC between the poles, now in the presence of different tunable parameters $\alpha$ and $\kappa$. For this analysis, we deliberately use structures that have properties of high absorption loss. For the corresponding values of $\alpha$, we find a different $\kappa$, which is other than the $\kappa$ for the quasi-BIC we have reported earlier. We report five of such quasi-BICs in the parameter ($\kappa,\alpha$) plane that have similar properties as reported above. Upon close inspection and using a curve fitting tool, we have discovered that all these special points, identified and labeled as quasi-BICs (using the similar process of ARC) lie on a straight line in the parameter  ($\kappa,\alpha$) plane. This straight line, which connects all the discrete quasi-BICs as depicted in fig 3(a), is referred to as a special-BIC line.  A simple technique to check if any point is a quasi-BIC in the parameter space is to check the ratio of the $Q$-factor of the resonance that shows quasi-BIC feature at the configuration defined by the $(\kappa,\alpha)$ parameters to the $Q$-factor of the same resonance for a passive cavity. The presence of a quasi-BIC would be well established when the ratio, as mentioned above, would diverge to enormous values. Furthermore, we find that the quasi-BIC at $\alpha=2.23$ and $\kappa=0.0527$ lie on this special-BIC line which we cross-verify by extrapolating the special-BIC line. Interestingly, this line could be instrumental for understanding the BIC-physics straightforwardly as the cumbersome task of identification and labeling of a quasi-BIC could be bypassed. Also, the formation of such a BIC-line introduces a new degree of freedom for the exploration of unconventional BIC-physics in various cavity geometries. 

For a more detailed study of these quasi-BICs, we study the isolated $Q$-factors of the same resonances under the different quasi-BIC configurations. The same has been plotted in fig 3(b), where the quality factor variation with $\kappa$ has been color codded with the five quasi-BICs, as can be seen in fig 3(b). We notice that the increase in $\alpha$ would require a lesser amount of $\kappa$ to be introduced in the system.  This inverse relationship between the parameters is of paramount importance from the fabrication point of view. With the help of state-of-the-art fabrication and implementation techniques, the novel scheme could be implemented with ease in different applications such as low-threshold lasers and integrated photonic devices for higher optical performance.

Furthermore, in the direction of understanding the device application and the utility of the special-BIC line, we define a new parameter $W=10log(Q/Q_0)$, where $Q$ quality factor at a given set of parameters $(\kappa,\alpha)$ and $Q_0$ is the $Q$-factor at the quasi-BIC. Here, we express $W$ in decibels. This parameter, $W$, captures the change in quality factor due to the change in the parameters of the system. This is interesting in the sense of detecting any small perturbation inducing material changes in the system. These small perturbations that make changes in the system could, as a consequence, change this sensitivity coefficient $W$. Since $Q$ is a quantity that is measurable with ease with modern advances and technological development, thereby parameter $W$ would unarguably be more physically realizable. Similar to the transmission sensitivity study we have performed, we try to see the sensitivity of quality factor as we move away from the quasi-BIC point of operation. Here, we plot sensitivity in terms of quality factor as opposed to sensitivity in terms of transmission characteristics. We perform the analysis of keeping the gain constant and varying the $\alpha$ parameter. We try to plot for all the five quasi-BICs we obtain, and curve fit them for better visualization and extrapolation. We observe more than 12.7 dB change in degree of confinement, when there is a 10\% change in $\alpha$ as compared to the configuration supporting quasi-BIC. In fact, BIC-2 achieves a sensitivity factor of more than 20dB with the same change. Upon much closer inspection, we find that the curve has significant change with even smaller changes away from the BIC. The magnitude of the change for smaller perturbation as small as 1\% for BIC-$i$ for $i={1,2,3,4,5}$ are as high as 9.6 dB for BIC-2. Please note that the nomenclature for naming the BICs is the same as expressed in the special-BIC line. Unlike transmission properties having an empirical dependence on the material properties, there seemed to be no distinctive concrete relationship between quality factors of different quasi-BICs. For illustration, BIC-2 has maximum sensitivity despite having lower $\kappa$ values or lower gain in the system to close the radiation channel. Therefore, we can safely say that gain hereby in such a system that is governed by ARC, would have the function of making sure that the leakage losses are being reduced to null, despite the fact that the system overall lossy.These results, in emphasis, could be implemented in different structures to enhance different optical performances; however, the underlying principle guiding the quasi-BIC would be similar.

In summary, we explore the phenomenon of ARC to enhance the degree of confinement of light in a Fabry-Perot type resonator equipped with varying gain and loss character. We notably show the ability to host a quasi-BIC through proper parameter tuning. We demonstrate how to exploit the presence of a quasi-BIC in terms of performance of a sensor in two ways-transmission and quality factor sensing. We theoretically study in depth the formation of a quasi-BIC and its consequences. We show a drastic divergence of a lifetime up to 4 orders of magnitude as we approach quasi-BIC state. By controlling these parameters, we show that it is possible to host multiple BICs in a cavity between the same pair of connected proximity resonances. This fine-tuning ability was extended where all these distinct multiple BICs led to the formation of a novel special-BIC line to provide a new degree of freedom to study the BIC-physics in resonators. These results to optimize modes to give ultra-enhanced optical performance opens up a massive potential in terms of many applications such as bio-sensing and imaging, high-performance integrated photonic devices, low threshold nano/micro-lasers, and device-level sensors. We believe that these results, through state-of-the-art fabrication techniques and with more advancements in resonator physics would open up further research and approach in various fields of non-linear optics and meta material physics.

%The authors acknowledge useful fruitful discussion with Govid P. Agarwal.
We acknowledge the financial support from the Science and Engineering research Board (SERB) [Grant No. ECR/2017/000491] and Ministry of Human Resource Development (MHRD), Government of India.


\begin{references} 
	\bibitem{Vahala}
K. Vahala, \href{https://doi.org/10.1038/nature01939}{Nature \textbf{424}, 839–846 (2003)}. 
\bibitem{Lin}
Z. Lin, X. Liang, M. Loncar, S.G. Johnson, and A.W. Rodriguez, \href{https://doi.org/10.1364/OPTICA.3.000233}{Optica \textbf{3}(3), 233 (2016)}.

\bibitem{biosensing}
T. Yoshie, L. Tang, and S.Y. Su, \href{https://doi.org/10.3390/s110201972}{Sensors 11(2), 1972 (2011)} 
\bibitem{Lasing}
W. Huang, Y.-H. Liu, K. Li, Y. Ye, D. Xiao, L. Chen, Z.-G. Zheng, and Y.J. Liu, \href{https://doi.org/10.1364/OE.27.010022}{Opt. Express \textbf{27}(7), 10022 (2019)}.	

\bibitem{Quantum}
T.-J. Wang, S.-Y. Song, and G.L. Long,
\href{https://doi.org/10.1103/PhysRevA.85.062311}{Phys. Rev. A \textbf{85}, 062311 (2012)}.	

\bibitem{Soljacic}
M. Soljacic, M. Ibanescu, S. G. Johnson, Y. Fink, and J. D.
Joannopoulos,\href{https://doi.org/10.1103/PhysRevE.66.055601}{Phys. Rev. E \textbf{66}, 055601(R) (2002)}.;
M. Soljacic, C. Luo, J. D. Joannopoulos, and S. Fan,\href{https://doi.org/10.1364/OL.28.000637}{Opt. Lett. \textbf{28}, 637 (2003)}.

\bibitem{yuri}
K. Koshelev, S. Lepeshov, M. Liu, A. Bogdanov, and Y. Kivshar,
\href{https://doi.org/10.1103/PhysRevLett.121.193903}{Phys. Rev. Lett. \textbf{121}, 193903 (2018)}.

\bibitem{Tolstoy}
I. Tolstoy, \href{https://doi.org/10.1121/1.395141}{J.Acoust. Soc. Am. \textbf{80}, 282 (1986)}; I. Tolstoy, \href{https://doi.org/10.1121/1.396389}{ J.Acoust. Soc. Am. \textbf{83}, 2086 (1988)}.

\bibitem{bic}
M.V. Rybin, K.L. Koshelev, Z.F. Sadrieva, K.B. Samusev, A.A. Bogdanov, M.F. Limonov, and Y.S. Kivshar,
\href{https://doi.org/10.1103/PhysRevLett.119.243901}{ Phys. Rev. Lett. {\bf 119}, 243901 (2017)}.

\bibitem{bic2}
A.A. Bogdanov, K.L. Koshelev, P.V. Kapitanova, M.V. Rybin, S.A. Gladyshev, Z.F. Sadrieva, K.B. Samusev, Y.S. Kivshar,
and M.F. Limonov, \href{https://doi.org/10.1117/1.AP.1.1.016001}{
	Advanced Photonics {\bf 1}, 016001 (2019)}.

\bibitem{bic3}
M. Rybin, and Y. Kivshar, \href{https://doi.org/10.1038/541164a}{Nature \textbf{541}, 164–165 (2017)}.

\bibitem{Hsu1}
C.W. Hsu, B. Zhen, A.D. Stone, J.D. Joannopoulos, and M. Soljacic, \href{https://www.nature.com/articles/natrevmats201648}{Nat. Rev. Mat. \textbf{1}, 16048 (2016)}.

\bibitem{Harsh1}
H.K. Gandhi, D. Rocco, L. Carletti and, C.D. Angelis, \href{https://doi.org/10.1364/OE.380280} {Opt. Exp. \textbf{28}(3), 3009 (2020)}.

\bibitem{Laha}
A.~Laha, and S.~Ghosh, \href{https://doi.org/10.1364/OL.41.000942}{Opt. Lett. \textbf{41}(5), 942 (2016)}; H.K. Gandhi, A.~Laha, and S.~Ghosh, \href{https://arxiv.org/abs/1905.13642} {arXiv:1905.13642 (2019)}.

\bibitem{ARC1}
J. Wiersig, \href{https://doi.org/10.1103/PhysRevLett.97.253901}{Phys. Rev. Lett. 97, 253901 (2006)}.



\bibitem{Song}
Q.H. Song, L. Ge, J. Wiersig,and H. Cao, \href{https://doi.org/10.1103/PhysRevA.88.023834}{Phys. Rev. A \textbf{88}, 023834 (2013)}; Q.H. Song,and H. Cao, \href{https://doi.org/10.1103/PhysRevLett.105.053902}{Phys. Rev. Lett. \textbf{105}, 053902 (2010)}.

\bibitem{Stillinger}
F.H. Stillinger,and D.R. Herrick, \href{https://doi.org/10.1103/PhysRevA.11.446}{Phys. Rev. A \textbf{11}, 446--454 (1975)}; K. Koshelev, A. Bogdanov, and Y. S. Kivshar, \href{https://doi.org/10.1016/j.scib.2018.12.003}{Sci. Bull. 64, 836 
	(2019)}.
%H.K. Gandhi, A.~Laha, and S.~Ghosh, \href{https://arxiv.org/abs/1905.13642} {arXiv:1905.13642 (2019)}.

\bibitem{BIC}
C.W. Hsu, B. Zhen,J. Lee, S.L. Chua, S.G. Johnson, J.D. Joannopoulos, and M. Soljacic, \href{https://www.nature.com/articles/nature12289}{Nature \textbf{499}, 188 (2013)}; F. Monticone, and A. Alu, \href{https://journals.aps.org/prl/pdf/10.1103/PhysRevLett.112.213903}{Phys. Rev. Lett. \textbf{112}, 213903 (2014)}.

\bibitem{Marinica}
D.C. Marinica, A.G. Borisov,and S.V. Shabanov, \href{https://doi.org/10.1103/PhysRevLett.100.183902}{Phys. Rev. Lett \textbf{100}, 183902 (2008)}.

\bibitem{Ordonez}
G. Ordonez, K. Na, and S. Kim, \href{https://journals.aps.org/pra/pdf/10.1103/PhysRevA.73.022113}{Phys. Reev. A \textbf{73}, 022113 (2006)}.

\bibitem{Carletti}
L. Carletti, K. Koshelev, C.D. Angelis, and Y.S. Kivshar, \href{https://doi.org/10.1103/PhysRevLett.121.033903}{Phys. Rev. Lett. \textbf{121}, 033903 (2018)}.

\bibitem{FW}
H. Friedrich and D.Wintgen, \href{https://doi.org/10.1103/PhysRevA.32.3231}{ Phys. Rev. A \textbf{32}, 3231 (1985)}.

\bibitem{jin}
J. Jin, X. Yin, L. Ni, M. Soljacic, B. Zhen, and C. Peng, \href{https://doi.org/10.1038/s41586-019-1664-7}{Nature \textbf{574}, 501 (2019)}.   

\bibitem{Carletti2}
L. Carletti, S.S. Kruk, A.A. Bogdanov, C.D. Angelis, and Y. Kivshar, \href{https://doi.org/10.1103/PhysRevResearch.1.023016}{Phys. Rev. Research 1, 023016 (2019)}.


\bibitem{Teller}
E. Teller, \href{https://doi.org/10.1021/j150379a010}{J.Phys. Chem. \textbf{41}(1), 109 (1937)}.

\bibitem{Phang}
S. Phang, A. Vukovic, S.C. Creagh, T.M. Benson, P.D. Sewell and G. Gradoni, \href{https://doi.org/10.1364/OE.23.011493}{Opt. Exp. \textbf{23}, 11493 (2015)}.


\bibitem{Ge}
L. Ge, Y. D. Chong, S. Rotter, H. E. Türeci, and A. D. Stone,
\href{https://doi.org/10.1103/PhysRevA.84.023820}{Phys. Rev. A 84, 023820 (2011)}.


\end{references}
\end{document}